# Subtle Sensing: Detecting Differences in the Flexibility of Virtually Simulated Molecular Objects

**Rhoslyn Roebuck Williams**
Intangible Realities Laboratory,
University of Bristol
Bristol, UK
rhoslyn.roebuckwilliams@bristol.ac.uk

**Xan Varcoe**
Intangible Realities Laboratory,
University of Bristol
Bristol, UK
xan@varcoe.com

**Becca R. Glowacki**
Goldsmiths College,
University of London
b.glowacki@gold.ac.uk

**Ella M. Gale**
Intangible Realities Laboratory,
University of Bristol
Bristol, UK
ella.gale@bristol.ac.uk

**Alexander Jamieson-Binnie**
Intangible Realities Laboratory,
University of Bristol
Bristol, UK
alexander.jamieson-binnie@bristol.ac.uk

**David R. Glowacki**
Intangible Realities Laboratory,
University of Bristol
Bristol, UK
glowacki@bristol.ac.uk

## Abstract
During VR demos we have performed over last few years, many participants (in the absence of any haptic feedback) have commented on their perceived ability to 'feel' differences between simulated molecular objects. The mechanisms for such 'feeling' are not entirely clear: observing from outside VR, one can see that there is nothing physical for participants to 'feel'. Here we outline exploratory user studies designed to evaluate the extent to which participants can distinguish quantitative differences in the flexibility of VR-simulated molecular objects. The results suggest that an individual's capacity to detect differences in molecular flexibility is enhanced when they can interact with and manipulate the molecules, as opposed to merely observing the same interaction. Building on these results, we intend to carry out further studies investigating humans' ability to sense quantitative properties of VR simulations without haptic technology.

## CSS Concepts
• **Human-centered computing~Human computer interaction (HCI)**; *Haptic devices*; User studies;

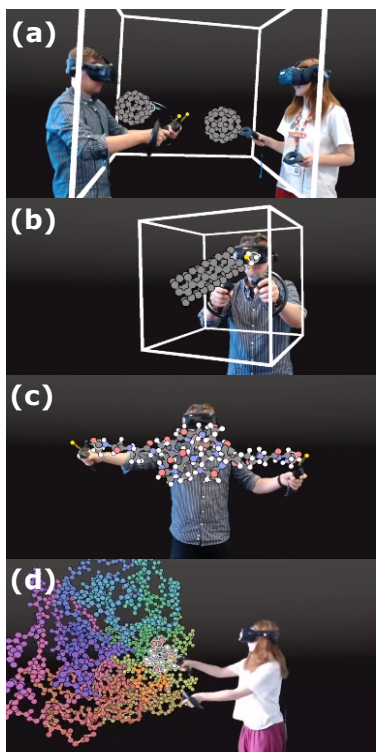

Figure 1: Manipulating molecular structures using Narupa: (a) $C_{60}$ molecules, (b) nanotube and methane, (c) a protein string, and (d) a drug bound to a protein. Hyperlinks to videos (a)–(d):
vimeo.com/382823399
vimeo.com/382823867
vimeo.com/382823683
vimeo.com/382824155

## Haptic Design in VR

Increasingly, virtual reality (VR) [16] and other immersive technologies are being used across a range of domains, including manufacturing [5,10], surgical training [1,13] and education [3,6]. This paper explores the application of VR to the domain of scientific visualization, and molecular simulation more specifically. Fred Brooks [4] and Kent Wilson [2] were among the first to explore whether immersive forms of HCI could be used to go beyond the rigid, physical 3d molecular models often found in laboratories and classrooms. They sought to create interactive, flexible and dynamic molecular models, whose atoms the user could touch and move, and whose physical and mechanical responses were simulated using rigorous laws of physics. Over the last few decades, increased computational power alongside technological advances in VR mean that molecules can be simulated at interactive speeds, a recent modelling paradigm that we call interactive molecular dynamics in VR (iMD-VR).

With the development of immersive simulation environments like VR, certain applications have sought to integrate haptic actuators to create experiences that allow participants to reach out and 'feel' force feedback from virtual objects. Surgical simulations are amongst the most successful scientific domains where haptics have been integrated with VR [1,9,17], e.g. replicating the resistance of human tissue against a surgical needle. For surgical simulators, where the intent is to mimic a well-defined 'real-world' experience as closely as possible, it is possible to carry out detailed comparisons of the extent to which the *virtually* simulated haptic forces mimic *real*-world forces, and the accuracy of the virtual experience can be carefully and iteratively refined. For applications like manipulating the nanoscale dynamics of flexible molecular objects such as those shown in Fig 1, there is no equivalent well-defined design reference. This is because molecular manipulation represents a class of experience for which there is no real-world sensory analogue for detailed comparisons of the real-to-virtual mapping. In such cases, it is far less clear how haptic technology should be applied.

Further difficulty often arises from the fact that haptics represent costly non-commodity pieces of equipment that require significant hardware, software, and driver maintenance. As such, their technological cost and sophistication can outweigh their benefits. Moreover, haptic technologies face fundamental limitations owing to the fact that there are no *generalized* solutions in the form of a single device that enables participants of a VR environment to feel *anything* (e.g., in the same way that visual or auditory display can be programmed to display anything). As a result, VR pioneers like Mel Slater argue that a generalized haptic solution is likely only possible in the form of a direct brain interface – i.e., as a form of applied neuroscience [12].

## "Subtle Sensing" in VR

Given the inherent difficulties of applying haptic technologies to domains like molecular simulation, we have begun to investigate the extent to which human sensory systems are able to detect haptic sensations in the absence of haptic actuators. Such sensing is more "subtle" than that which occurs with haptic motors. It is grounded in somatic (bodily) sensitivity, proprioception (our non-visual understanding of the position of our body in space), and careful attention to movement. Such 'subtle-sensing' methods have already been observed to accentuate participants' somatic awareness – e.g., in the somaesthetic approach recently discussed by Höök *et al.* [8]. The ability of subtle methods to alter perception has similarly been highlighted in the field of *pseudo*-haptics. For example, Pusch *et al.* [14] investigated creating the perception of force feedback in VR by breaking proprioception. Users placed their hand into a virtual force field and attempted to keep it in the same position by counteracting the force, after which most participants reported feeling a 'pushing' sensation on their hands. Other examples combine the breaking of proprioception with vibrational feedback to give both tactile feedback and emulate kinesthetic feedback (e.g., Rietzler *et al.* [15]).

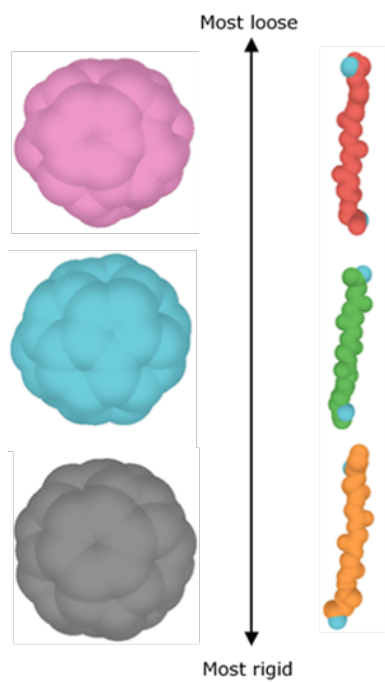

Figure 2: Showing the three virtual $C_{60}$ molecules (left) and strings (right), in order of rigidity.

**"Subtle Sensing" of Molecular Properties**
Over the last several years, we have developed an open source software framework called Narupa [18], a VR simulation environment which allows multiple users to co-habit the same virtual environment and interact with molecules in real-time, which is finding use in both scientific research [12] and education [3]. Introducing haptic tools to simulate force feedback is difficult not only for the reasons mentioned above, but also because Narupa tends to be used with multiple participants in the same virtual space, as shown in Fig 1a. For this reason, we have begun to explore the more subtle aspects of perception highlighted in the pseudo-haptics and somatics literature. During demonstrations performed in our lab over the last couple years, many participants report 'feeling' differences between molecules, despite the lack of haptic actuators. Interestingly, participants consistently give similar descriptions of each type of molecule. For example, participants often describe the respective buckminster fullerene ($C_{60}$) and nanotube molecules shown in Figs 1a and 1b as 'rigid' and the protein shown in Fig 1c as 'soft' or 'floppy'.

The ability to carry out 'subtle sensing' by tuning into more subtle aspects of our perception through pseudo-haptic approaches (e.g., to detect the flexibility or rigidity of simulated molecular structures) would allow participants to carry out scientific tasks in VR more effectively. This will have utility in both educational and scientific contexts because it furnishes insight into the function, design, and interactions that characterize molecules. One such example is drug design (Fig 1d), where the participant must gently place the drug into the active site of a protein, whilst only minimally perturbing the protein's structure. The more effectively a human can sense subtle interactions between the drug and protein, the more likely they are to find viable low-energy binding pathways and design better drugs.

**Pilot User Study Design**
In what follows, we present preliminary work we have undertaken to understand the extent to which participants could correctly detect differences in the flexibility and rigidity of simulated virtual molecular objects. We used virtual molecules to carry out this study for three main reasons: (1) it is straightforward to tune molecular flexibility and rigidity using the physics simulation engine available within Narupa; (2) molecules are unfamiliar to most people, thereby avoiding familiarity bias; and (3) there is a large range of molecules available for use in Narupa.

Each experiment involved two participants, each wearing an HTC-Vive headset. As shown in Fig 3, one participant (the 'handler') was given a set of handheld HTC Vive VR controllers, and the other participant (the 'observer') was not. We compared the ability of both participants to correctly rank the mechanical properties of three virtual molecular objects within the virtual environment. The objects differed only in their elasticity/rigidity and their colors (Fig 2). Tests were performed for two types of molecules, whose physics were simulated in real-time using Narupa.

Table 1: Showing the coefficients of the angle force constants of the bonds in the three $C_{60}$ molecules.

| Rigidity | Force constant coefficient |
|---|---|
| Least | 1 |
| Middle | 3 |
| Most | 9 |

Table 2: Showing the coefficients of the bond force constants of the bonds in the backbone of the three polypeptide strings.

| Stretchiness | Force constant coefficient |
|---|---|
| Least | 3 |
| Middle | 1 |
| Most | 1/3 |

Since we wished to understand whether participants could interpret the mechanical properties of the virtual objects, it made sense to use ones that had some resemblance to real-world objects. We settled on two simulations: (1) a 'ball' shaped $C_{60}$ molecule (or 'buckyball') and (2) a short, 'string' shaped polypeptide.

The next challenge was to decide which mechanical properties of the virtual objects to change. We wished to emulate changes in the *softness* or *rigidity* of the $C_{60}$ molecules, and the *stretchiness* or *elasticity* of the strings, since these are intuitive descriptions of mechanical properties that have some mapping onto real-world balls and strings. Narupa models interatomic bonds and angles as anharmonic springs, each characterised by a *force constant*. Modification of these force constants causes mechanical changes in molecular flexibility and rigidity. We experimented with a range of force constant modifications to evaluate how different modifications were sensed by the participants. For the $C_{60}$, it became clear that modifying the *angle* force constants allowed us to change how easy it was for the participant to deform the 'roundness' of the buckyball. For the strings, the *bond* force constants caused a change in perceived elasticity. We decided to only change the force constants of the bonds in the

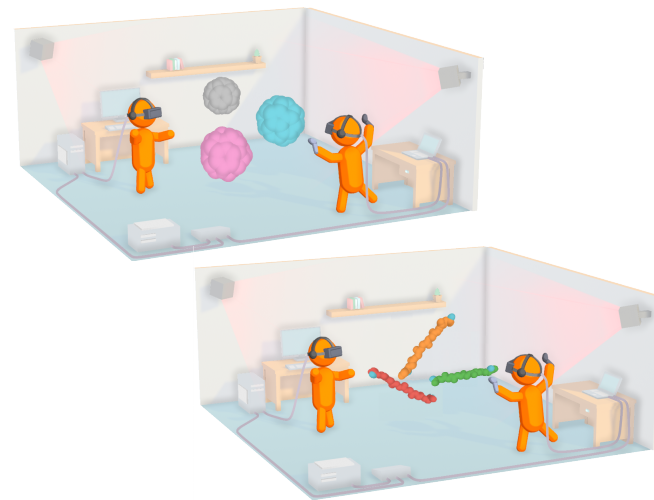

Figure 3: the experimental setup with two participants (an observer on the left and a handler on the right) with three $C_{60}$ molecules (top) and with three strings (bottom).

*backbone* of the polypeptide string, since the backbone defines the main structure of the string.

### Determining the force constant values

To decide which values of the force constant coefficients should be used during our pilot user study, we ran a small, iterative exploratory study. This study was run with one handler and one observer cohabiting the Narupa environment, whilst we iteratively adjusted the force constants of each of the three objects (either $C_{60}$ molecules or strings) based on live feedback from the participants. For each pair of observers, once the observer indicated *that they could not sense a difference* between the flexibility/rigidity of the molecules and the handler indicated *that they could sense a difference*, we recorded the corresponding set of force constant values for each of the three $C_{60}$ molecules and strings [Table 1 and Table 2]. We then took an average over all the sets of force constants and chose these for our pilot user study. The representation of the objects was identical apart from the colours, which were varied as shown in Fig 2, in order that the participants could distinguish between different molecules.

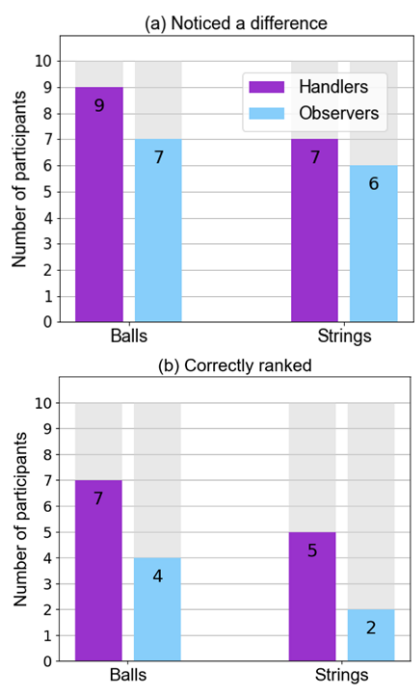

Figure 4: For the $C_{60}$ molecules and strings separately, the number of handlers (10 total) and observers (10 total) who (a) noticed a difference between the three objects, and (b) correctly ranked all three objects.

## Pilot User Study: Experimental Details

Having determined the force constants, we ran a pilot user study using a separate cohort (N=20). Seven participants had some experience in VR, whilst five had moderate experience. Of the remaining participants, two had no experience and two considered themselves very experienced. Most participants had a degree in higher education, and many were affiliated with the University of Bristol. There were approximately equal numbers of men (n=11) and women (n=9).

In each experiment, both participants were first presented with a nanotube simulation, enabling them to familiarize themselves with molecular manipulation in Narupa. They then transitioned into a second environment, which contained each of the three $C_{60}$ molecules (shown in Fig 2 and 3). Taking note of the colors, the observer was invited to watch as the following instructions were directed to the handler:

1. Hold onto either side/end of each object in turn, and move it around in front of you. Feel free to play with it as you wish.
2. Gently "pull the molecules apart" by moving your hands apart and then bringing them close together.
3. Push the two ends of each object towards one another by bringing your hands close together, then bring your hands apart again.

The participants then exited VR and were presented with pictures of the three differently-colored virtual objects. They were asked to write down any observations they had about the objects. Once finished, they were invited back into a VR simulation containing the three colored strings, where they were taken through the same sequence as before. Upon finishing the tasks, they were taken out of VR and asked to write down any observations they had about the strings. Finally, they were provided with a questionnaire that asked: (1) Did you notice a difference in the elasticity of any of the $C_{60}$ molecules? (2) Did you notice a difference in the elasticity of any of the three strings? (3) If so, for both object types please rank them in terms of their elasticities. For each question, pictures of the differently colored objects were provided. Each study was performed by reading off a script to control the information provided to the participants. Although sections of the script were directed at either the handler or observer specifically, for each experiment both participants were present in the same room and heard the entirety of the script. The force constants were kept the same across the experiments.

## Results and Discussion

The results show that more handlers than observers noticed a difference between the three objects for both the $C_{60}$ molecules and strings (Fig 4a). In addition, more handlers also *correctly* ranked the elasticities of the three objects than the observers (Fig 4b). Taking

Table 3: Showing the p-values calculated using Fisher's exact test.

|  | **Fisher's test p-value** | |
| --- | --- | --- |
|  | Balls | Strings |
| Noticed a difference | 0.248 | 0.325 |
| Correct ranking | 0.150 | 0.146 |

**Acknowledgements**
DRG recognizes funding from the Leverhulme Trust (Philip Leverhulme Prize) and Royal Society (URF/R/180033); AJB and RRW are supported by the EPSRC TMCS CDT (EP/L015722/1); Ella M. Gale is supported by EPSRC grant EP/S024107/1. Support and insight at various stages came from Alex Jones, Michael B. O'Connor, Lisa Thomas, Helen Deeks, Oussama Metatla, and Thomas Mitchell.

the null hypotheses to be that all participants were equally capable of (1) noticing a difference between the three objects and (2) correctly ranking the three objects, we used the Fisher's exact test to calculate significance (Table 3). Although these results are not below a significant value of p=0.05, owing to the very small number of participants, there were consistently more correct answers from the handlers than the observers. Using a binary outcome superiority trial [19] on the number of correct rankings, we determined the number of participants required in future studies to achieve a typical significance level (0.05) and statistical power (0.8). Assuming similar results to those observed in the relatively small sample results shown in Fig 4, our analysis suggests that the $C_{60}$ molecules require 36 user studies (N=72), and the strings require 40 user studies (N=80). These will be used as a rough estimate for the number of participants required for future studies. However, these numbers may require some adjustment pending changes to the experimental protocol as we refine our methods.

Qualitatively, some interesting observations arose which we believe will inform future studies. When asked to describe their interaction with the $C_{60}$ molecules, one handler noted that the "pink [ball, which was the most elastic] felt 'fat' and 'heavy' whilst the grey [ball, which was the most rigid] had a higher frequency, quicker feel to it". When asked about their interaction with the strings, the same participant said "it felt that the chains had different lengths rather than elasticity". This participant correctly ranked the buckyballs, however, answered that all three strings had identical elasticities. This indicates that more participants than recorded may have noticed differences between the objects but that the study and questionaire design did not accurately capture their impressions. A number of participants also indicated to us that, upon exiting VR to fill out the questionnaire and rank the objects' relative elasiticies, they were unable to remember 'which color was which'. Moving forward, we intend to design the study so as to reduce questionnaire ambiguity and also to minimize potential memory effects.

**Conclusions and Further work**
This study presents preliminary data suggesting that VR participants may be able to utilize 'subtle sensing' methods to distinguish the mechanical properties of virtual molecular objects. The fact that the observers (who could rely only on vision) were unable to make determinations that were as accurate as the handlers (who could rely on vision, movement, and proprioception) suggests that 'subtle sensing' in VR may involve mechanisms that go beyond the purely visual. Several improvements will be made to future studies, including: (1) randomizing the object colors to diminish potential association of mechanical features with colors, and (2) randomizing the order in which the objects are presented to participants to account for potential practice and/or learning effects [11] and varying ability to recall information presented at the beginning, end, and middle of the experiment. We also wish to gauge the quantitative limit of subtle sensing mechanisms – i.e., to quantitatively determine whether there is a point at which users are no longer able to resolve differences. We also intend to explore whether handlers wearing recently developed VR gloves [7] influences their ability to correctly rank the mechanical properties of the objects. This further work will enable us to better understand the degree to which 'subtle sensing' mechanisms might be extended to other domains of scientific simulation and visualization.